\documentclass[twocolumn,aps,floatfix]{revtex4}

\usepackage{mathrsfs,amsmath,amssymb,bm}
\usepackage{lipsum}
\usepackage{amsfonts}
\usepackage{graphicx}
\usepackage{epstopdf}
\usepackage{makecell, rotating}
\usepackage{multirow}
\usepackage{algorithmic}
\usepackage{multirow}
\usepackage{array, diagbox}
\usepackage{color}
\usepackage{epsfig,subfig,dcolumn}

\usepackage[draft]{changes}
\definechangesauthor[name={Kai Jiang}, color=red]{jk}
\newcommand{\br}{\bm{r}}

\usepackage{enumerate}
\usepackage{ulem}%下划线
\usepackage{setspace}
\usepackage{epstopdf}

\begin{document}

\title{Phase behavior of symmetric diblock copolymers under 3D soft confinement}

\author{ Zhijuan He$^{1\dagger}$, {Jin Huang}$^{2,1\dagger}$, Kai Jiang$^{1*}$ and An-Chang Shi$^{3*}$}

\affiliation{
$^{1}$School of Mathematics and Computational Science, Xiangtan University,  
Hunan Key Laboratory for Computation and Simulation in Science and Engineering, and Key Laboratory of Intelligent Computing and Information Processing of Ministry of Education, 411105, P.R. China \\
$^{2}$Laboratory of Mathematics and Complex Systems (Ministry of Education), 
	School of Mathematical Sciences, 
	Beijing Normal University, Beijing, 100875, P.R. China\\
$^{3}$Department of Physics and Astronomy, McMaster University, 
	Hamilton, Ontario L8S 4M1, Canada}

\thanks{Authors to whom correspondence should be addressed. 
	E-mail: kaijiang@xtu.edu.cn (K. Jiang), shi@mcmaster.ca (A.-C. Shi).\\
$\dagger$~ These authors contributed equally to this work.
}

\begin{abstract}
The phase behavior of symmetric diblock copolymers under three-dimensional (3D) soft confinement 
is investigated using the self-consistent field theory. 
The soft confinement is realized in binary blends composed AB diblock copolymers and C homopolymers, where the copolymers self-assemble to form a droplet embedded in the homopolymer matrix. The phase behavior of the confined block copolymers is regulated by the degree of confinement and the selectivity of the homopolymers, resulting in a rich variety of novel structures. 
When the C homopolymers are neutral to the  A- and B-blocks, stacked lamellae (SL) are formed where the number of layers increases with the droplet volume, resulting in a morphological transition sequence from Janus particle to square SL. 
When the C homopolymers are strongly selective to the B-blocks, a series of non-lamellar morphologies, including onion-, hamburger-, cross-, ring-, and cookie-like structures, are observed. 
A detailed free energy analysis reveals a first-order reversible transformation between SL and onion-like (OL) structures when the selectivity of the homopolymers is changed.
Our results provide a comprehensive understanding of how various factors, such as the copolymer concentration, homopolymer chain length, degree of confinement, homopolymer selectivity, affect the self-assembled structures of diblock copolymers under soft 3D confinement.
\end{abstract}

\maketitle

\section{Introduction}
\label{sec:intrd}
Block copolymers (BCPs) have been attracting sustained attention due to their ability to self-assemble into numerous ordered nanostructures\,\cite{Hamley2003Nanostructure, bates2017block}. 
These ordered nanostructures have wide applications in various fields, such as catalysis, electronics, biomedical engineering, and optical sensors\,\cite{mai2012self}.
The phase behavior of BCPs in bulk and under confinement has been widely studied experimentally\,\cite{nan2018recent,deng2021responsive, hu2023structure}
and theoretically\,\cite{zhang2017research, zhang2018confined,hiroshi2014frustrated,huang2021block,shi2013selfassembly}.
These previous studies clearly established that the self-assembly of BCPs under confinement can lead to the formation of a variety of morphologies that are not available in unconfined systems, providing a powerful technique for fabricating novel ordered nanostructures.

The self-assembly of BCPs is largely driven by the intrinsic frustration of the BCPs\,\cite{Shi2021frustration}. 
Placing BCPs under confinement introduces additional factors, such as geometric frustration, interfacial interactions, symmetry breaking, and entropy loss, that will affect their self-assembly behavior. 
The nature of confinement could be classified by the geometrical dimension of the confining environment as one-dimensional (1D) (thin film), two-dimensional (2D) (nanotube or cylinder), and three-dimensional (3D) (sphere or ellipse).
In particular, the phase behaviors of AB diblock copolymers under 1D and 2D confinement have been studied extensively\,\cite{zhang2017research,  zhang2018confined, shi2013selfassembly}.
Under 1D confinement or in thin films, the self-assembled morphology primarily depends on the film thickness\,\cite{Lambooy1994Observed, Koneripalli1996Confinement, WangQ2000Symmetric, WangQ2001Monte, Yin2004Simulated} and the surface preference of to the different blocks
\cite{Kellogg1996Observed, Yokoyama2000Structure,
	Hao2017Self, WangQ2000Symmetric, WangQ2001Monte, Yin2004Simulated,
	Vu2011Controlling}. 
Under 2D confinement where BCPs are placed in a cylindrical space, the self-assembled morphology is significantly influenced by the surface field\,\cite{Sevink2001Morphology, Chen2006Effect}
and the pore diameter\,\cite{Xiang2004Block, Xiang2005The, Wu2004Composite,
	He2001Self, Chen2006Effect}. 

Under 3D confinement that could be realized by placing BCPs in spherical cavities\,\cite{Arsenault2005Block,wu2023nanospheres} or conical nanopores\,\cite{Kim2019Self}, the self-assembled morphologies depends strongly on the size and surface preference of the confining environment\,\cite{nan2018recent,deng2021responsive,hu2023structure,zhang2017research,zhang2018confined,guo2024fabrication,huang2024design,yan2023confined,zheng2021halogen,xu2015abctriblock,Jeon2007Nanostructures,li2020kinetically}.
%\cite{  nan2018recent, zheng2021halogen, deng2021responsive, Khandpur1995Polyisoprene, Bate1994Fluctuations,Jeon2007Nanostructures, Arsenault2005Block,Higuchi2008Frustrated,xu2015abctriblock,deng2015soft, Kim2019Self,cui2019janus,li2020kinetically,mao2021polymersome, yan2023confined,ghosh2019nanoparticles,deng2013reversible, Ku2019Shape,wu2022emulsion,charlotte2011interplay, zheng2021halogen,liu2023asymmetric}
%Moreover, theoretical studies can offer comprehensive insights into the phase behavior of self-assembling BCPs in confinement. Theoretical calculations have explored the stability of different structures under specific conditions, including the degree of polymerization (molecular weight), volume fraction, degree of confinement, and the affinity of each block to the confining surface.
Monte Carlo simulations\,\cite{BinYu2007Self,BinYu2011Confined, wu2023nanospheres,he2018self} and self-consistent field theory (SCFT)\,\cite{PengChen2008Microstructures, zhao2021aysmmetric} have been used to study the self-assembly of diblock copolymers confined in spherical and ellipsoidal cavities. 
These previous theoretical and simulation studies revealed that the self-assembled morphologies and chain conformations of BCPs under 3D confinement depend sensitively on the degree of confinement characterized by the cavity size and the surface interactions with the blocks. 
More interestingly, the studies predicted that 3D confinement could induce the formation of various complex nanostructures such as \mbox{toroidal,} \mbox{single helical,} \mbox{cylindrical helical,} saddle-related and nanospheres with patches in polyhedral structures. 
It is noted that the aforementioned studies are carried out for hard confinement, where the geometry of the confining environment is fixed. In contrast, in many cases soft confinements are realized where the confining environment is not fixed. In this case the self-assembly of BCPs is accompanied by the deformation of the confining environment, resulting in richer phase behavior.

Due to their rich phase behaviors, the self-assembly of BCPs under soft confinement has attracted significant attention in recent years. One simple system to realize 3D soft confinement of BCPs is obtained in the blends containing BCPs and homopolymers. Under appropriate conditions, macroscopic phase separation of the blends would lead to the formation of BCP-rich droplets suspended in the homopolymer-rich matrix. Several experimental studies of the self-assembled morphologies of diblock copolymer droplets have been carried out\,\cite{reffner1992thesis, Satio2007Effect, Jeon2007Nanostructures, Tanaka2009Control, deng2013reversible, zheng2021halogen,liu2023asymmetric,Higuchi2008Spontaneous,Higuchi2010Phase, Higuchi2013Reorientation,kim2012golddecorated}.
It has been observed that the simultaneous BCP self-assembly and droplet deformation can enrich the phase behavior of the system. 
For example, Reffner\,\cite{reffner1992thesis} observed that BCPs droplets could self-assemble into numerous frustrated structures, such as concentric shells of packed cylinders, concentric shells of lamellae, and layers of honeycomb structures.
Moreover, the self-assembly of diblock copolymers under emulsion solvent evaporation-induced conditions was studied by Zhu and coworkers\,\cite{zheng2021halogen,liu2023asymmetric, deng2013reversible}. They observed various special nanostructures, such as pupa-like, bud-like, onion-like (OL), Janus-type, honeycomb-like structures, stacked toroids in prolate ellipsoids and inverse curved cylinders in ellipsoids. 
Furthermore, the phase-separated structures of diblock copolymer nanoparticles, as investigated by Higuchi et al.\,\cite{Higuchi2008Spontaneous,Higuchi2010Phase, Higuchi2013Reorientation}, exhibited a variety of ordered nanostructures, including OL, stacked lamellae (SL), Janus-type, tennis-ball-, mushroom-, wheel-, or screw-like structures.  
Theoretically, simulated annealing method has been used to study the phase behaviors of diblock copolymers in poor solvents with different bulk phases\,\cite{Peng2011Soft} or  homopolymers\,\cite{zhang2017soft}. A series of phases including single-helix, double helices, one-side-connected double helices, stacked toroids, pine-cone-like, pupa-like structures were predicted. 
Monte Carlo simulation method has been employed\,\cite{Yan2016Self} to investigate the self-assembled morphologies of AB-type diblock copolymers confined in a 3D soft nanodroplet. 
They observed a series of phase transitions induced by the confinement degree. 
SCFT has been utilized to study the self-assembly of AB  diblock copolymers in C homopolymers\,\cite{xia2014self} or solution with different molecular size\,\cite{ wang2011polymer}.
They observed the layer, vesicle, and micelle phases.
Dynamic self-consistent field theory\,\cite{zhu2023process} has been used to explore the kinetic path of the formation of BCP particles.
They explained the formation mechanism of striped ellipsoids, OL particles and double-spiral lamellar particles.
Despite these previous theoretical studies, the formation mechanism of many novel nanostructures in 3D soft confining BCPs remains incomplete.

\begin{figure}[htbp]
	\centering
	\begin{minipage}[t]{1.0\linewidth}
		\centerline{\includegraphics[scale=1.0]{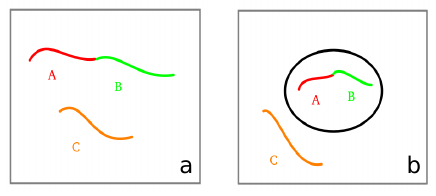}}
	\end{minipage}
	\caption{
		(a) Schematic of the binary blend composed of AB diblock copolymers and C homopolymers. 
		(b) Schematic of a BCP-rich droplet embedded in the homopolymer-rich matrix. 
		%Color scheme: A (red), B (green), C (orange), and the interface of AB diblock copolymers and C homopolymers (black).
	}
	\label{fig: AB_C_shown}
\end{figure}

In this paper, we report a theoretical study on the self-assembly of AB diblock copolymers under 3D soft confinement by using the highly successful polymeric self-consistent field theory (SCFT). The soft confinement of BCPs is realized in a binary blend  containing AB diblock copolymers and C homopolymers, as illustrated in FIG.\,\ref{fig: AB_C_shown}.
The interaction parameters of the system are adjusted so that the binary blend undergoes macroscopic phase separation between the AB diblock copolymers and the C homopolymers, resulting in the the formation of copolymer-rich droplets embedded in the homopolymer-rich matrix. The BCP droplet is then used as a model system to study the self-assembly of BCPs under 3D confinement. 
Our results indicate that the phase behavior of the BCP droplet is significantly influenced by the concentration of AB diblock copolymers, the chain length of C homopolymers, the degree of confinement, and the strength of homopolymer selectivity. As the concentration of AB diblock copolymers increases or the degree of confinement decreases, the number of layers in the SL increases. Moreover, a reversible first-order transition between the SL and OL structures can be observed by adjusting C homopolymers preference for B-monomers.

\section{Theoretical framework}
\label{sec:theory}

The binary blend model used in our study is composed of $n_{\text{AB}}$ AB diblock copolymers and $n_{\text{C}}$ C homopolymers in a volume $V$. 
Each copolymer chain contains two blocks, {\it i.e.} an A-block with $\text{N}_{\text{A}}$ A-monomers and a B-block $\text{N}_{\text{B}}$ B-monomers. The total degree of polymerization of the copolymer is 
$\text{N} = \text{N}_{\text{A}} + \text{N}_{\text{B}}$, and the volume fraction of the $\alpha$-type monomers is $f_{\alpha} = \text{N}_{\alpha}/\text{N}$, where $\alpha \in \{ \text{A}, \text{B} \}$. 
Each C homopolymer chain is characterized by a degree of polymerization $\text{N}_{\text{C}}$ or a normalized chain length $\kappa = \text{N}_{\text{C}}/\text{N}$. The volume of each monomer $\rho_{0}$, or equivalently the monomer density $\rho_{0}=1/v_0$, is assumed to be the same, so that the concentration or volume fraction of the AB diblock copolymers in the system is given by $\text{c} = n_{\text{AB}} \text{N}/(n_{\text{AB}} \text{N} + n_{\text{C}} \text{N}_{\text{C}})$. 
%Here, we use the Gaussian chains\,\cite{Fredrickson2006The} to model the system.  
Furthermore, that A-, B-, and C-blocks are assumed to be Gaussian chains with the same statistic length $b$. 

Within the mean-field approximation the free energy of the blend in unit of $k_{\text{B}} T$, where $k_{\text{B}}$ is the Boltzmann constant and $T$ is a given temperature, can be expressed as,\,\cite{Fredrickson2006The}
\begin{align}\label{eq:ziyouneng}
	\frac{\text{NF}}{Vk_BT}&=
	\frac{1}{V} \int \sum_{\alpha \neq \beta}  
	\chi_{\alpha \beta} \text{N} \rho_{\alpha}(\bm{r}) \rho_{\beta}(\bm{r})\,d\bm{r} 
	\notag \\
	& -\! \frac{1}{V} \int 
	\sum_{\alpha} w_{\alpha}(\bm{r}) \rho_{\alpha}(\bm{r})\,d \bm{r} 
	\! +\! \frac{1}{V}\int  w_{+}(\bm{r})\left(\rho_{+} (\bm{r})\! -\! 1\right)\,d\bm{r}
	\notag \\
	&-\text{c} \ln Q_{\text{AB}}
	-(1-\text{c}) \ln Q_{\text{C}},
\end{align}
where $\chi_{\alpha \beta}$ is the Flory-Huggins interaction parameter characterizing the interaction between the $\alpha$- and $\beta$-monomers ($\alpha,\beta \in \{ \text{A}, \text{B}, \text{C}\}$). The fields $\rho_{\alpha}(\br)$ and $w_{\alpha}(\br)$ are the segment density and conjugate field of the $\alpha$-type monomers, respectively. The function $w_+(\br)$ is a Lagrange multiplier used to enforce the incompressibility condition $\rho_{+}(\br)=\rho_{A}(\br)+\rho_{B}(\br)+\rho_{C}(\br)=1$. 
The two quantities $Q_{AB}$ and $Q_{C}$ are the single chain partition functions of AB diblock copolymers and C homopolymers, respectively, which are obtained from the following expressions,
\begin{align}
	& Q_{\text{AB}} =
	\frac{1}{V} \int  q_{K}(\bm{r}, s) q^{\dagger}_{K}(\bm{r},f_{K}-s)\,d \bm{r}, 
	\, \forall \, s\in [0,f_{K}],\notag \\
	& Q_{\text{C}} = 
	\frac{1}{V} \int q_{\text{C}}(\bm{r}, s) q^{\dagger}_{\text{C}}(\bm{r},\kappa -s)\,d \bm{r},
	\, \forall \, s \in [0,\kappa] \notag.
\end{align}
Here, the chain propagators $q_{\alpha}(\bm{r},s)$ and 
$q^{\dagger}_{\alpha}(\bm{r},s)$, $ \alpha \in \{ \text{A}, \text{B}, \text{C}\}$, 
represent the probability of finding the $s$-th segment at the spatial position $\bm{r}$. 
These chain propagators are obtained as solutions of the modified diffusion equations (MDEs),
\begin{equation}\label{eq:chuanbozi}
	\begin{aligned}
		\frac{\partial}{\partial s} q_{\alpha}(\bm{r}, s) 
		& = R_g^2 \nabla^{2} q_{\alpha}(\bm{r}, s) 
		- w_{\alpha}(\bm{r}) q_{\alpha}(\bm{r}, s), 
		\\ 
		-\frac{\partial}{\partial {s}} q^{\dagger}_{\alpha}(\bm{r}, s) 
		& = R_g^2 \nabla^{2} q^{\dagger}_{\alpha}(\bm{r}, s) 
		- w_{\alpha}(\bm{r}) q^{\dagger}_{\alpha}(\bm{r}, s),
	\end{aligned}
\end{equation}
with the initial conditions,
$q_{\text{A}}(\bm{r},0) = q_{\text{C}}(\bm{r},0) = q^{\dagger}_{\text{B}}(\bm{r},f_{B})= q^{\dagger}_{\text{C}}(\bm{r},\kappa) = 1$, 
%$q_{\text{B}}(\bm{r},0) = q_{\text{A}}(\bm{r},f_{\text{A}})$, 
%	$q^{\dagger}_{\text{A}}(\bm{r},0) = q^{\dagger}_{\text{B}}(\bm{r},f_{\text{B}})$. 
Here $R_g$ is the gyration radius of an ideal Gaussian chain. 

The mean-field the SCFT equations are obtained by demanding that the first-order variations of the free energy functional with respect to densities and mean fields vanish,
$\delta \text{F}/\delta \rho_{\alpha} = \delta \text{F}/ \delta w_{\alpha} = \delta \text{F}/\delta w_{+} = 0$, leading to the following SCFT equations,
\begin{small}
	\begin{align}
		w_{\text{A}}(\bm{r})\! &=
		\chi_{\text{AB}} \text{N} {\rho}_{\text{B}}({\bm{r}}) + \chi_{\text{AC}} \text{N} \rho_{\text{C}}({\bm{r}}) + w_{+}({\bm{r}}), \\
		w_{\text{B}}({\bm{r}})\! &=
		\chi_{\text{AB}} \text{N} \rho_{\text{A}}({\bm{r}}) + \chi_{\text{BC}} \text{N} \rho_{\text{C}}({\bm{r}}) + w_{+}({\bm{r}}) ,\\
		w_{\text{C}}({\bm{r}})\! &=
		\chi_{\text{AC}} \text{N} \rho_{\text{A}}({\bm{r}}) + \chi_{\text{BC}} \text{N} \rho_{\text{B}}({\bm{r}}) + w_{+}({\bm{r}}), \\
		1\! &=\rho_{\text{A}}({\bm{r}})+\rho_{\text{B}}({\bm{r}})+\rho_{\text{C}}({\bm{r}}),\\
		\rho_{\text{A}}({\bm{r}}) \!  &=\!
		\frac{\text{c}}{Q_{\text{AB}}}  
		\int_{0}^{f_{\text{A}}}  q_{\text{A}}({\bm{r}}, s) q^{\dagger}_{\text{A}}({\bm{r}}, f_{\text{A}}\!-\!s)\, ds, 
		\label{eq:solve_rho_1}\\ 
		\rho_{\text{B}}({\bm{r}}) \!   &= \!
		\frac{\text{c}}{Q_{\text{AB}}}
		\int_{0}^{f_{\text{B}}}   q_{\text{B}}({\bm{r}}, s) q^{\dagger}_{\text{B}}({\bm{r}}, f_{\text{B}}\!-\!s)\, ds,
		\label{eq:solve_rho_2}\\ 
		\rho_{\text{C}} ({\bm{r}}) \!  &= \!
		\frac{1-\text{c}}{Q_{\text{C}}}  
		\int_{0}^{\kappa}  q_{\text{C}}({\bm{r}}, s)  q_{\text{C}}^{\dagger}(\bm{r}, \kappa\!-\!s)\, ds. 
		\label{eq:solve_rho_3}
	\end{align}
\end{small}

The SCFT equations are a set of nonlinear and nonlocal equations with multiple solutions. Each solution of the SCFT equations correspond to a local minimum, or a possible phase, of the system. Finding the solutions of the SCFT equations is a major task of the study. The primary procedure of solving the SCFT equations is as follows:
\begin{enumerate}[Step 1]
	\item\label{Step_initial} Specify the initial mean fields $w_{\alpha}(\bm{r})$, $\alpha \in \{ \text{A}, \text{B}, \text{C}\}$. 
	\item\label{Step_propagators} Obtain the propagators 
	$q_{\alpha}(\bm{r},s)$, $q^{\dagger}_{\alpha}(\bm{r},s)$ 
	by solving the MDEs\,\eqref{eq:chuanbozi}.
	\item\label{Step_densities} Compute the densities $\rho_{\alpha}(\bm{r})$ by using the integral equations\,\eqref{eq:solve_rho_1}-\eqref{eq:solve_rho_3}.
	\item\label{Step_F} Calculate the free energy F.
	\item\label{Step_fields} Update the mean fields $w_{\alpha}(\bm{r})$ and goto Step \ref{Step_propagators} if the free energy difference between two consecutive iterations is larger than a given error $\epsilon$.
\end{enumerate}

In this work, we assume periodic boundary conditions and use the pseudo-spectral method \cite{tzeremes2002efficient, rasmussen2002improved,  jiang2010spectral} to treat the spatial discretization. 
We apply the fourth-order Adams-Bashford scheme to discrete the MDEs in the $s$-direction
\,\cite{Cochran2006Stability}. 
The four initial values of the fourth-order Adams-Bashforth scheme are obtained using the Richardson's extrapolation method\,\cite{Ranjan2008Linear} based on the second-order operator-splitting method\,\cite{rasmussen2002improved}. 
The integral equations \eqref{eq:solve_rho_1}-\eqref{eq:solve_rho_3} are solved by the composite Simpson's formula with fourth-order truncation error\,\cite{Press2007The}.
Due to its local convergence, we first take the simple mixing iteration method to obtain better initial values for mean fields \cite{Drolet1999Combinatorial}, and then use the Anderson mixing algorithm \cite{thompson2004improved} to accelerate the convergent process.
In the Anderson mixing algorithm, we update the mean fields by using information from previous $70$ steps.

\section{Results and discussion}
\label{sec:results}
All spatial functions in our computations are expanded in terms of plane waves. The size of the cubic computational box is $[0,L]^3$ wiht $L=64$ and $64 \times 64 \times 64$ plane-wave basis functions are used to discretize the 3D unit cell.
The step along the chain contour length and the degree of polymerization are taken as $\Delta s=0.0025$ and $\text{N} = 100$, respectively.
To study the phase behavior of symmetric AB diblock copolymers and C homopolymers, we fix the volume fraction of the A-blocks at $f_{\text{A}} = f_{\text{B}} = 0.5$.
The SCFT computations are carried out until the free energy difference $\epsilon$ between two consecutive steps is less than $ 1 \times 10^{-8}$.
Owing to the significant macroscopic phase separation between the symmetric AB diblock copolymers and the C homopolymers, all figures below display only the density distribution of the A- and B-monomers to enhance the visibility of morphology changes. The A- and B-rich domains are plotted in red and green, respectively.

\begin{figure*}[htbp]
	\centering
	\begin{minipage}[t]{0.1\linewidth}
		\centerline{\includegraphics[scale=0.12]{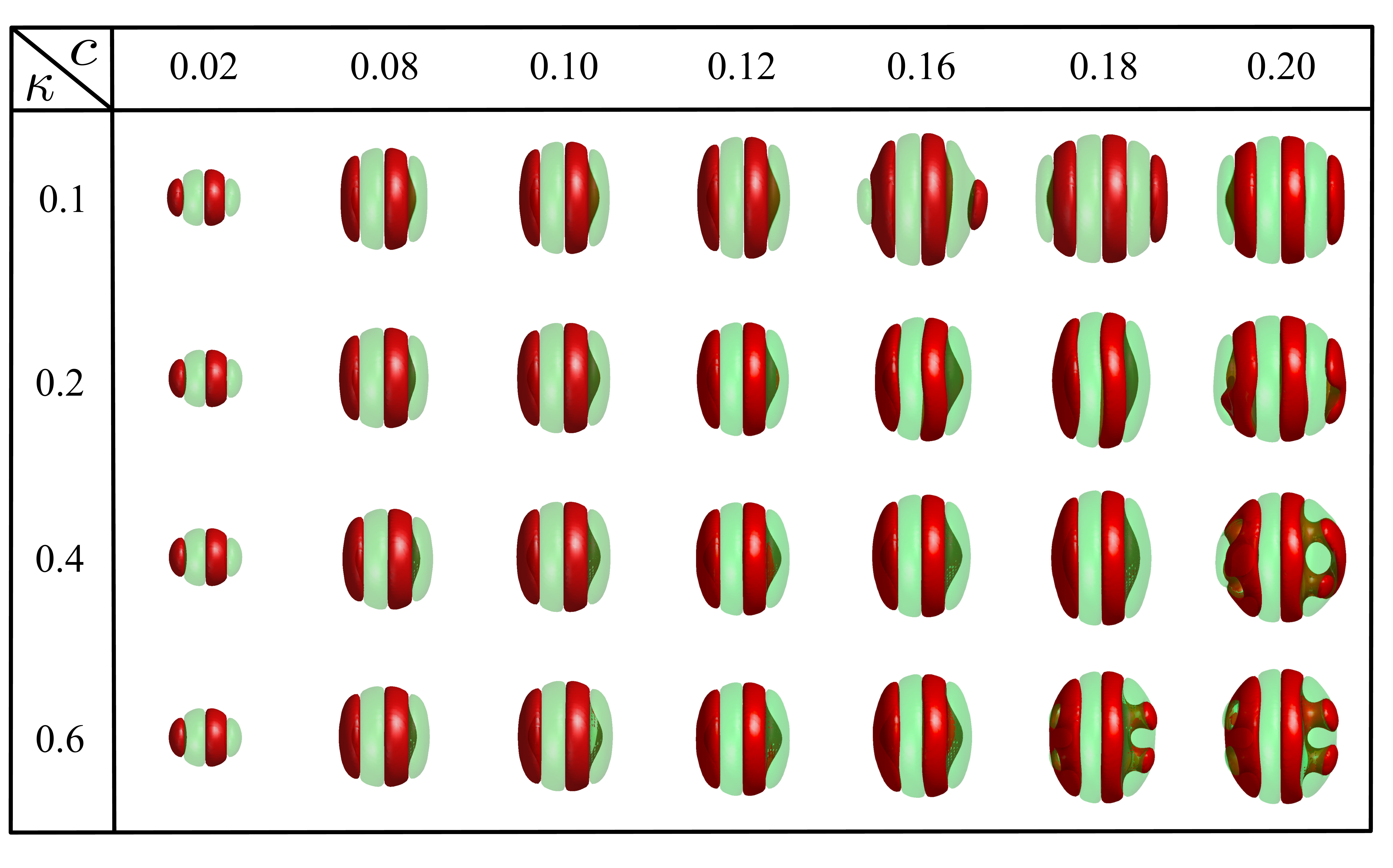}}
	\end{minipage}
	\caption{Morphologies of symmetric AB diblock copolymers under soft confinement as a function of $c$ and $\kappa$ with the following parameters,
		$[\chi_{\text{AB}} \text{N}, \chi_{\text{AC}}\text{N}, \chi_{\text{BC}}\text{N}]=[24, 24, 24]$, $L=15 R_g$. 
		To enhance the visibility of morphology changes, only the density distributions of A- and B-monomers are shown. The red, green represent the A-, B-rich domains, respectively.
	}
	\label{fig:gamma_c}
	
\end{figure*}

%---explain of fC and c 
\subsection{Effect of copolymer concentration and homopolymer chain length}

\begin{figure}[htbp]
	\centering
	\begin{minipage}[t]{1.0\linewidth}
		\centerline{\includegraphics[scale=1.5]{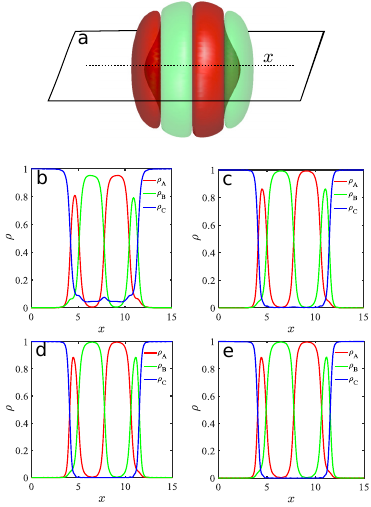}}
	\end{minipage}
	\caption{(a) The profile of SL4 along a symmetrical line  
		for $c=0.1$ when parameters 
		$[\chi_{\text{AB}} \text{N}, \chi_{\text{AC}} \text{N}, \chi_{\text{BC}} \text{N}]=[24, 24, 24]$, 
		$\text{c}=0.1$, $L=15\,R_{g}$ (see FIG.\,\ref{fig:gamma_c}), $x \in [0, L]$.
		$x$ is the variable along the symmetrical line.
		(b)-(e) show the density distributions of monomers as a function of $x$ with
		(b) $\kappa = 0.1$; (c) $\kappa = 0.2$; 
		(d) $\kappa = 0.4$; (e) $\kappa = 0.6$.
		The red, green, and blue lines represent the density distributions of A-, B-, and C-monomers along $x$, respectively.
	}
	\label{fig:change_fC}
\end{figure}  

\begin{figure}[htbp]
	\centering
	\begin{minipage}[t]{0.28\linewidth}
		\centerline{\includegraphics[scale=0.28]{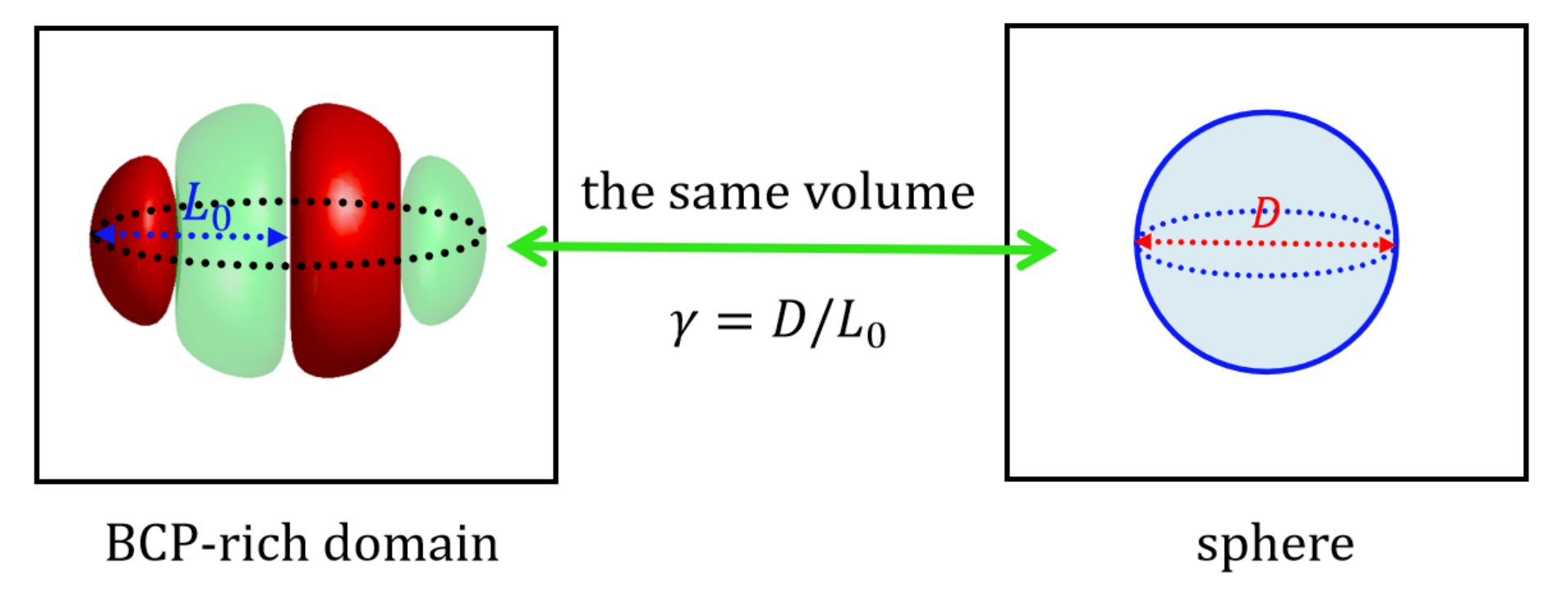}}
	\end{minipage}
	\caption{Schematic diagram of the definition of $\gamma$ for BCP-rich domain.}
	\label{fig:gamma}
\end{figure} 	

We first examine the influence of the concentration c of the BCPs and the normalized chain length $\kappa$ of the homopolymers on the self-assembled morphologies with $\chi_{\text{AB}} \text{N} = \chi_{\text{AC}}\text{N} = \chi_{\text{BC}}\text{N} =24$, thus the homopolymer is neutral to the A and B blocks. The resulting morphologies, mainly SL, are summarized in FIG.\,\ref{fig:gamma_c}.
When $\kappa$ is fixed, increasing $\text{c}$ enlarges the volume of the copolymer droplet, resulting in an increase in the number of layers, from 4 to 6, in the SL.
For simplicity, we abbreviate the SL with four layers as SL4.
Moreover, when $\text{c}\le 0.12$, the value of $\kappa$ has little effect on the number of layers. Consistently, the copolymers all self-assemble into the SL4.
To understand the influence of $\kappa$ on morphologies, 
we plot the density distributions of SL4 along a symmetrical line at $\text{c} = 0.1$ (see FIG.\,\ref{fig:gamma_c}), where $x$ represents the symmetrical line, as shown in FIG.\,\ref{fig:change_fC} (a).
The density distributions of A-, B-, C-monomers as a function of 
$x$ with different $\kappa$ values are shown in FIG.\,\ref{fig:change_fC} (b)-(d). 
From these density distributions, we observe that the contact area between the diblock copolymers and the homopolymers reduces as $\kappa$ increases. The findings indicate that C homopolymers with shorter chain length or smaller $\kappa$ penetrate into the AB-rich domain more easily. This penetration may lead to an increase of the effective volume of the AB-rich domain, which would influence the degree of confinement in the system.

\begin{figure}[htbp]
	\centering
	\begin{minipage}[t]{0.33\linewidth}
		\centerline{\includegraphics[scale=0.33]{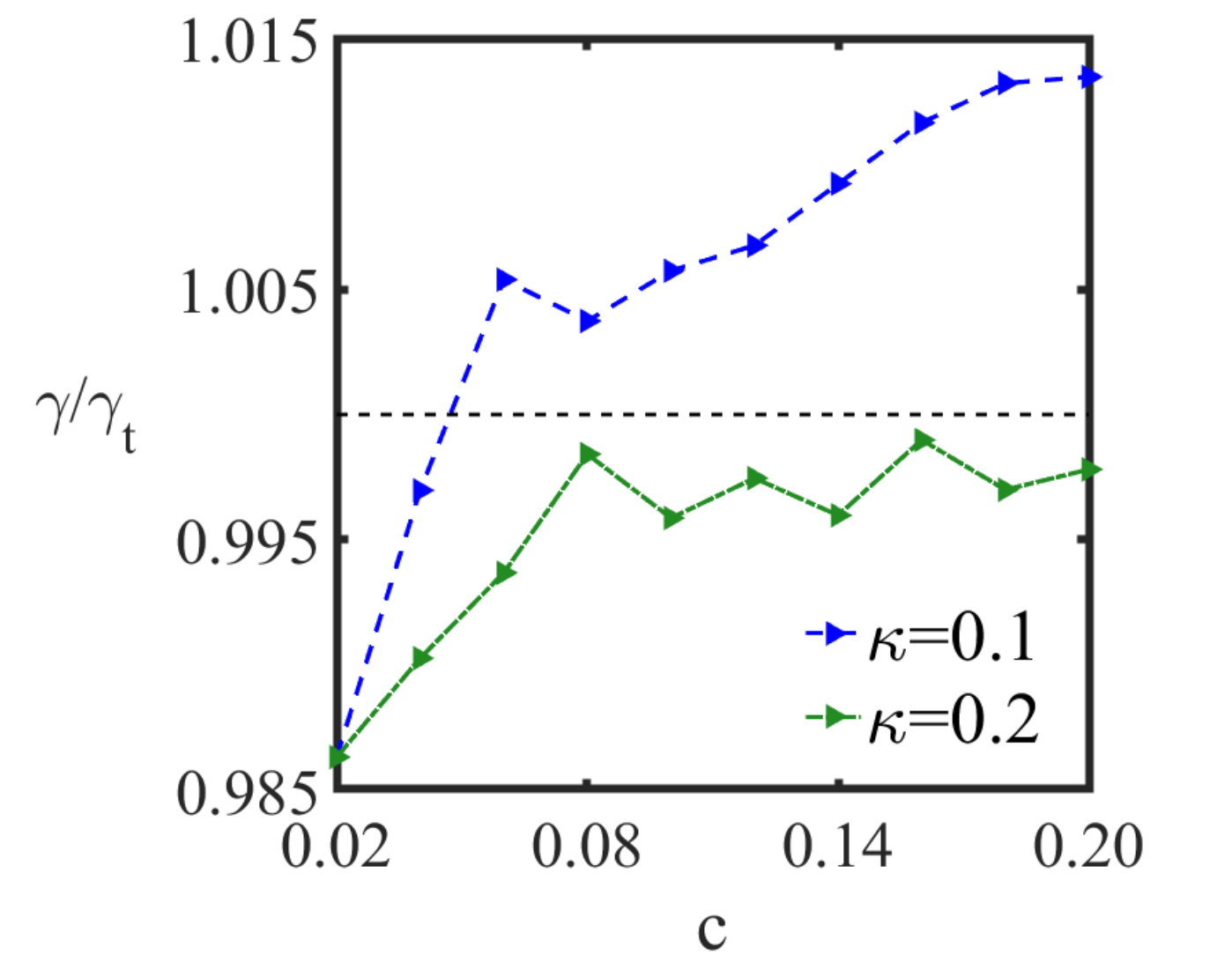}}
	\end{minipage}
	\caption{Ratio $\gamma/\gamma_t$ as a function of c with  $\kappa = 0.1$ (blue) and $\kappa = 0.2$ (green).
		Confinement degree	$\gamma$ is numerical calculated from the data in FIG.\,\ref{fig:gamma_c}, where $\gamma_t = 4.37c^{1/3}$ is obtained by theoretical derivation.}
	\label{fig:c_gamma_function}
\end{figure}

In order to better describe the effects of confinement in this system, we use the parameter $\gamma = D/L_0$ to quantify the degree of confinement of the BCP-rich domain, where $D$ is the diameter of a sphere with the same volume as the BCP-rich droplet (see FIG.\,\ref{fig:gamma}). 
The period of the SL, $L_0 = 4.26 R_g$, is obtained from the bulk symmetric AB diblock copolymers.
Since ${\text{c}} = (4\pi(D/2)^3/3)/L^3 = \pi D^3/{(6L^3)}$, we can obtain a theoretical value of $\gamma_{t} \approx 4.37 {\text{c}}^{1/3}$.
%Intuitively, $\gamma$ is positively correlated with c. The relationship between $\gamma$ and c can be also derived analytically. Note that ${\text{c}} = (4\pi(D/2)^3/3)/L^3 = \pi D^3/{(6L^3)}$, we can obtain $\gamma \approx 4.37 {\text{c}}^{1/3}$.
To further verify the relationship between $\gamma_{t}$ and c, we calculate the ratio $\gamma/\gamma_{t}$, %as a function of c with different $\kappa$ values, 
where $\gamma$ is numerically calculated from the data in FIG.\,\ref{fig:gamma_c}.
We take $\kappa = 0.1$ and $\kappa = 0.2$ as examples to better illustrate the comparison of the ratio $\gamma/\gamma_{t}$ obtained theoretically and numerically.
As shown in FIG.\,\ref{fig:c_gamma_function}, the resulting two lines fluctuate around the value 1, indicating that the numerical results are essentially consistent with the theoretical estimate.
The slight deviation between $\gamma$ and $\gamma_t$ may be because the BCP-rich domains are not entirely spherical and short C homopolymers penetrate into the BCP-rich domains, resulting in an overestimation of the BCP-rich domains volume.
Furthermore, when $\kappa=0.2$, the ratio $\gamma/\gamma_{t}$ is closer to the value of 1 than that of $\kappa=0.1$, possibly due to the fact that shorter C homopolymers more easily penetrate the BCP-rich domains, as shown in FIG.\,\ref{fig:change_fC} (b) and (c). 

%%%%%%%----------------------------------------------------------------------------------------
%%%%%%%------------------soft-confinement------------------------------------------------------
\subsection{Effect of the degree of confinement and the homopolymers selectivity}
\begin{figure*}[htbp]
	\centering
	\begin{minipage}[t]{0.1\linewidth}
		\centerline{\includegraphics[scale=0.08]{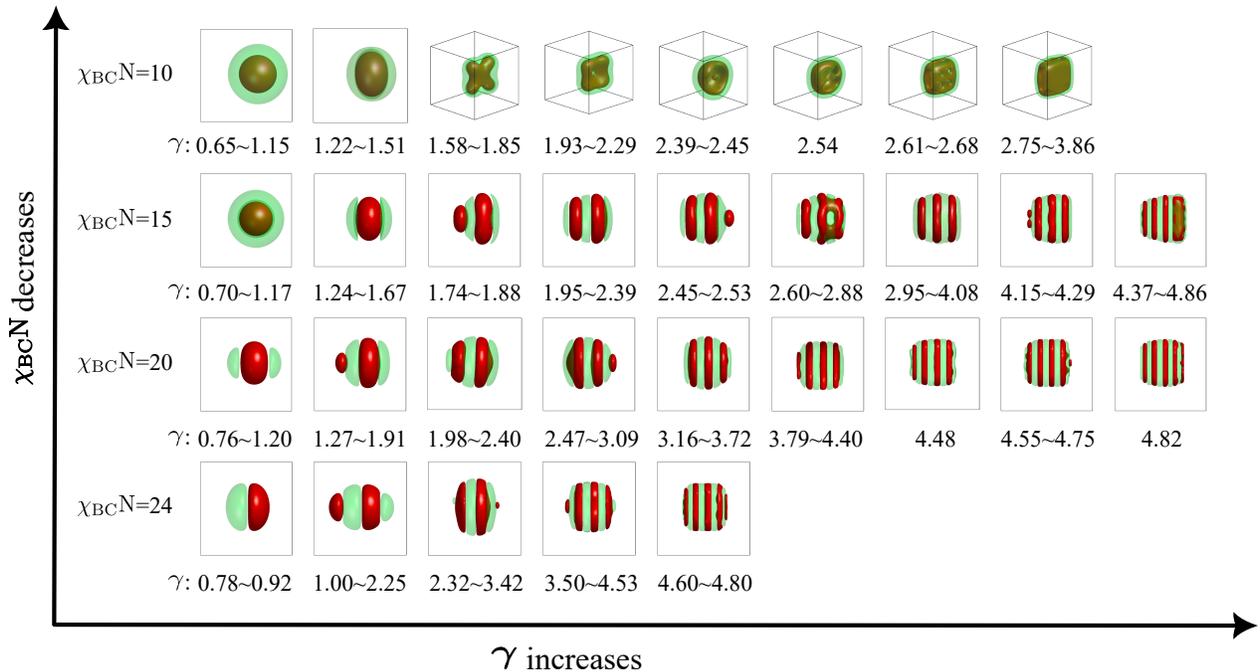}}
	\end{minipage}
	\caption{Morphologies of symmetric AB diblock copolymers 
		as a function of $\gamma$ and C homopolymers preference 
		when parameters 
		$[\chi_{\text{AB}} \text{N}, \chi_{\text{AC}} \text{N}]=[24, 24]$.
		The red, green colors represent the A-, B-rich domains, respectively.
	}
	\label{fig:soft_hard}
\end{figure*}

Here, we examine the influence of the degree of confinement $\gamma$ and the homopolymers selectivity on the self-assembly phase behaviors of the symmetric diblock copolymers under 3D soft confinement.
Three types of C homopolymer selectivities are considered: 
(a) neutral homopolymers with the same or similar AC and BC interactions, {\it i.e.} $\chi_{\text{AC}}\text{N}/\chi_{\text{BC}}\text{N}\approx 1$;
(b) B-selective homopolymer with weaker BC repulsive interaction, \textit{i.e.}, $\chi_{\text{AC}}\text{N}/\chi_{\text{BC}}\text{N} > 1$;
(c) strongly B-selective homopolymers with $\chi_{\text{AC}}\text{N}/\chi_{\text{BC}}\text{N} > \! > 1$.
In what follows we take $\chi_{\text{AB}}\text{N} = \chi_{\text{AC}}\text{N} = 24$,  $\text{c} = \kappa = 0.1$ and change the computational box size $L$ from $6 R_{g}$ to $35 R_{g}$ with a step size of $0.5 R_{g}$ to adjust $\gamma$.
As shown in FIG.\,\ref{fig:soft_hard}, we find that as the B-selectivity increases, the SL could transform into an OL structure and also self-assemble into many novel non-spherical structures. Moreover, as $\gamma$ increases, the number of layers in the SL increases.  The specific details are analyzed as follows.

For neutral homopolyers, $\chi_{\text{BC}} \text{N} = 24$, we observe a series of SL with different number of layers, all of which are composed of alternating A- and B-layers. 
The formation of SL can be attributed to the fact that the C homopolymers have the same preference for A- and B- monomers, as well as the fluidity of the copolymer/homopolymer interface.
Interestingly, as $\gamma$ increases, the system transitions from Janus particle to square SL, and the number of stacked layers gradually increases. This indicates that increasing the number of layers in SL is an effective strategy to accommodate the change in the degree of confinement. 
When the value of $\gamma$ increases from $1.00$ to $2.32$, from $2.32$ to $3.50$ and from $3.50$ to $4.60$, two additional A- and B-layers are formed in SL. This result indicates that under neutral conditions, the repeat period for SL is approximately $1.2L_0$.
This phenomenon has also been observed in experiments and theory\,\cite{Yan2016Self}.

For B-selective homopolymers with $\chi_{\text{BC}} \text{N} = 20$, the symmetric AB diblock copolymers also tend to assemble into the SL. The number of layers in SL increases progressively with the increase of the $\gamma$. It is evident that most of the surface area of the outmost A-layer is significantly smaller than that of B-layer (see FIG.\,\ref{fig:soft_hard}), causing SL to be asymmetric.
Moreover, SL with even layers have a wider range of $\gamma$ than those with odd layers.
Conversely, as the B-selectivity increases to $\chi_{\text{BC}} \text{N} = 15$, SL with odd layers have a wider range of $\gamma$ than those with even layers. 
The primary reason for this phenomenon may be that as the B-selectivity increases, C homopolymers tend to be in contact with B-monomers, thus making SL with two outmost B-layers more stable.

For strongly B-selective homopolymers with $\chi_{\text{BC}} \text{N} = 10$, 
we observe a series of complex morphologies, such as
onion- ($0.65<\gamma<1.15$), hamburger- ($1.22<\gamma<1.51$), cross- ($1.58<\gamma<1.85$), 
ring- ($1.93<\gamma<2.29$), and cookie-like ($2.61<\gamma<3.86$) structures. 
The characteristic feature of these structures is that A-monomers are enveloped by B-monomers. This formation may be primarily due to the fact that the C homopolymers have a stronger attraction to B-monomers than to A-monomers.

From the analysis above, it is evident that both $\gamma$  and the value of $\chi_{\text{BC}} \text{N}$  have significant effects on the self-assembled morphology of the BCP droplet.  
Interestingly, decreasing $\chi_{\text{BC}} \text{N}$, a transition from SL to OL structures 
is observed. 
Moreover, as $\gamma$ increases, the number of layers in SL increases, which can mitigate the impact of B-selectivity. Furthermore, we also discover that in SL, the size of the outermost layer is always smaller than that of the inner layers. Some similar phenomena have been also found in the spherical hard confinement system\,\cite{BinYu2007Self}. 
However, there exist differences of phase behaviors under 3D hard and soft confinement, largely due to the fact that the shape of the confining environment can change according to the self-assembled structure.
The changeable copolymer/homopolymer interface in soft confinement makes the structure less frustrated than that in hard confinement, providing more possibilities for the self-assembly phase behaviors.
Thus, various non-spherical structures, such as \mbox{hamburger-,} \mbox{cross-,} \mbox{ring-,} and cookie-like structures, can be obtained under soft confinement (see FIG.\,\ref{fig:soft_hard}).

\subsection{Transition between stacked lamellae and onion-like structures}

In the above sections, we have examined the effects of the degree of confinement $\gamma$ and B-selectivity ($\chi_{\text{BC}} \text{N}$) on the self-assembled morphologies. It was observed that the SL can transform into OL structures under the strongly B-selective condition. To further investigate this structural transition, we examine the behavior of the system by varying $\rm \chi_{BC} \text{N}$ with a set of fixed parameters $\rm \chi_{AB} \text{N}=24$, $\rm \chi_{AC} \text{N}=24$, $\rm L=7R_{g}$, $\rm c=0.09$, $\rm \kappa=0.25$. In the experiments carried out by Higuchi {\it et al.}\,\cite{Higuchi2010Phase, Higuchi2013Reorientation}, it was observed that at lower temperatures, disordered structures are formed in the nanoparticles. As the temperature rises, the disordered structures transition into SL. Further increasing temperature, the SL evolve into OL structures. 
To make our numerical computations consistent with the experimental conditions, we first use SL4 as the initial condition at $\chi_{\text{BC}} \text{N} = 24$. We simulate the condition of increasing temperature by carrying out SCFT calculations with $\chi_{\text{BC}} \text{N}$ gradually decreases in a step of $0.5$. During the simulations, the converged result of the current $\chi_{\text{BC}} \text{N}$ is used as initial condition for the next simulation with a decreased $\chi_{\text{BC}} \text{N}$. 
From these simulations, we observe a phase transition sequence from the SL4 to SL3, then to a dumbbell-like (DL) structure, and finally form an OL structure, as $\chi_{\text{BC}}\text{N}$ decreases.
We take these four structures as candidate phases and vary $\chi_{BC} \text{N}\in [12,24]$ to analyze this phase transition. 

\begin{figure}[htbp]
	\centering
	\begin{minipage}[t]{1.\linewidth}
		\centerline{\includegraphics[scale=0.23]{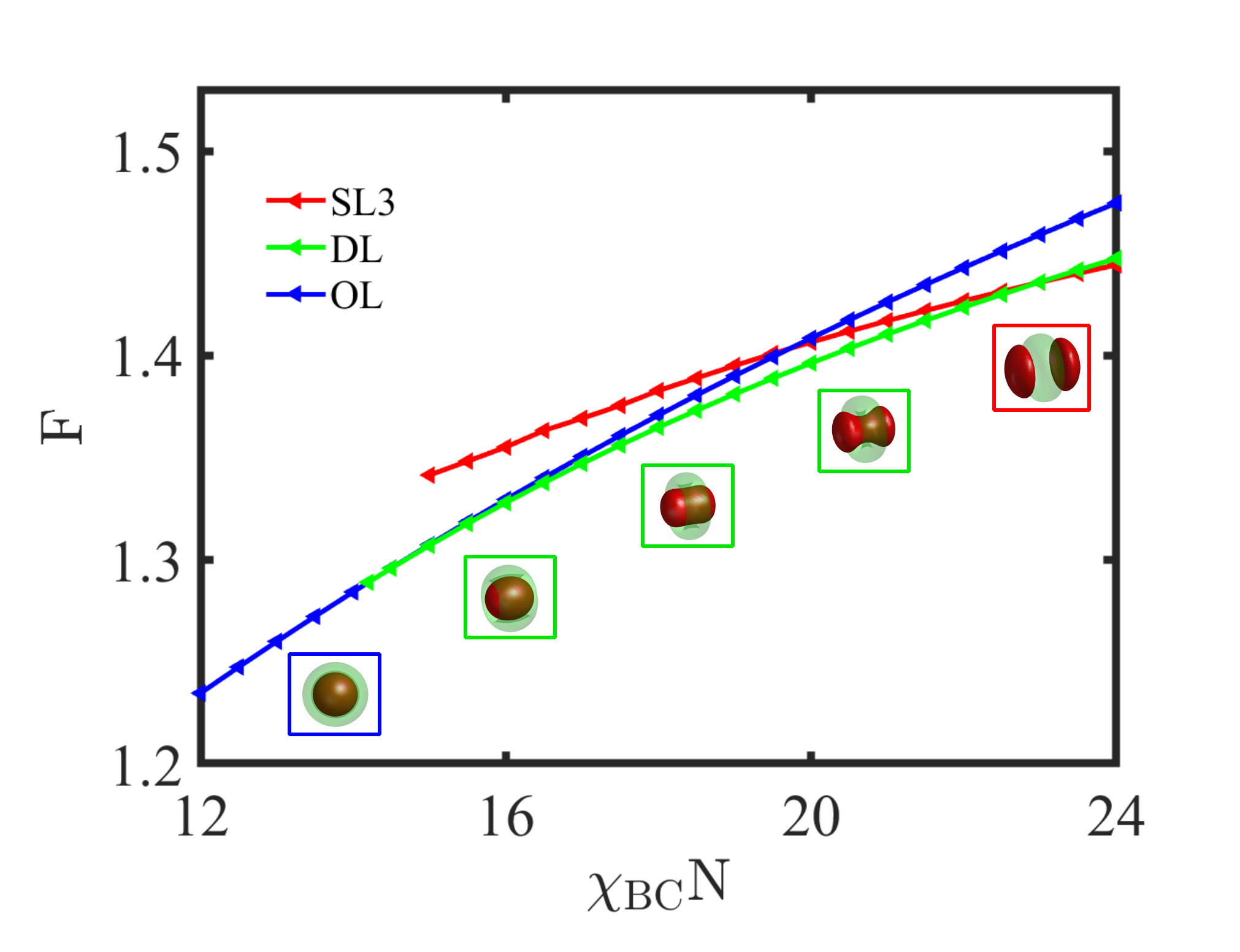}}
	\end{minipage}
	\caption{The free energy curves of SL3, DL and OL structures as a function of $\chi_{\text{BC}}\text{N}$ when 
		$\chi_{\text{AB}}\text{N}=24$, $\chi_{\text{AC}}\text{N}=24$, $L=7R_g$, $\text{c}=0.09$, $\kappa = 0.25$. 
	}
	\label{fig:OtoL}
	\label{fig:LtoO}
\end{figure}

Since the SL4 is metastable under theses parameters, we do not include it in the following free energy analysis.
As shown in FIG.\,\ref{fig:LtoO}, the phase transition from SL to OL structures can be verified as a first-order one. As the B-selectivity increases or the value of $\chi_{\text{BC}} \text{N}$ decreases, the A-monomers in the SL gradually aggregate and the contact area between the B-monomers and C homopolymers expands. This causes a phase transition from the SL3 ($22.95 \le \chi_{\text{BC}} \text{N} \le 24$) to DL structures ($14.3\le \chi_{\text{BC}} \text{N}< 22.95$), and eventually to the OL structures ($12 \le  \chi_{\text{BC}} \text{N} <14.3$). 
This phenomenon indicates that the increased B-selectivity of the decreased $\chi_{\text{BC}} \text{N}$ drives the C homopolymers to enlarge their contact with the B-monomers, resulting in B-monomers forming the outermost surface of the OL structure.
This mechanism explains the SL-OL transformation as $\chi_{\text{BC}} \text{N}$ decreases, which is consistent with with the experimental observations by Higuchi {\it et al.}\,\cite{Higuchi2010Phase, Higuchi2013Reorientation}.
To explore the reversibility of this transition, we set $\chi_{\text{BC}}\text{N} =12$ and use the OL structure as the initial condition for the first calculation. Then the initial condition of the next simulation is taken from the previously converged solution with $\chi_{\text{BC}} \text{N}$ gradually increasing in steps of $0.5$. 
We observe the phase transition from an OL structure to a DL structure, and finally to the SL3 as $\chi_{\text{BC}} \text{N} $ increases. 
Similarly, these three structures is used as candidate phases within the range of $ 12 \le \chi_{\text{BC}} \text{N} \le 24$, as shown in FIG.\,\ref{fig:LtoO}.
This result indicates that the phase transition between SL and OL structures is reversible.

\section*{Conclusions}
\label{sec:summary}
We systematically investigated the self-assembled morphologies of symmetric AB diblock copolymers under 3D soft confinement by using extensive SCFT calculations. We found that the copolymer concentration c, degree of confinement $\gamma$, and the C homopolymer selectivity play a dominate role in the formation of confined structures. 
As c increases, both the number of layers in SL and the surface area of the BCP-droplet increase.
For neutral homopolymers, the morphologies are SL with high symmetry and periodicity. The symmetry and periodicity of SL gradually weaken when the selectivity of the homopolymer increases. Under the strongly B-selective conditions, many novel morphologies are observed, including onion-, hamburger-, cross-, ring-, and cookie-like structures. 
Furthermore, a reversible first-order transformation between SL and OL structures is confirmed when the B-selectivity of the homopolymers or the value of $\chi_{\text{BC}} \text{N}$ is varied.
These results indicate that the binary blend of AB diblock copolymers and C homopolymers can be used as 
a model system for studying the self-assembly of diblock copolymers under soft confinement.
Further, the novel nanostructures predicted in this study could provide a guideline for the fabrication of structured nanoparticles using BCPs.
In the future, we intend to investigate the 3D soft confinement of various macromolecules, 
such as triblock copolymers, bottlebrushes and rigid rods. 
This direction would explore the possibility of obtaining structured nanoparticles with desirable internal morphologies and functions, which could be useful in potential applications in functional devices, drug delivery, and other emerging applications\,\cite{Qiang_2019_Template, Steinhaus_2018_Self, Staff_2012_Preparation, Deng_2012_Mesoporous, Zhang_2019_Enantiocomplementary}.

\renewcommand\refname{Reference}
\bibliography{rsc}

\end{document}